\documentclass[conference]{IEEEtran}

\IEEEoverridecommandlockouts
\usepackage{cite}
\usepackage{amsmath,amssymb,amsfonts}
\usepackage{algorithmic}
\usepackage{graphicx}
\usepackage{textcomp}
\usepackage{xcolor}
\usepackage{fancyhdr}
\usepackage{url}
\usepackage{hyperref}

\def\BibTeX{{\rm B\kern-.05em{\sc i\kern-.025em b}\kern-.08em
    T\kern-.1667em\lower.7ex\hbox{E}\kern-.125emX}}
    
\fancypagestyle{firstpagefooter}{%
  \fancyhf{}
  
  \fancyfoot[R]{}
}

\begin{document}

\title{Retrieval‑Augmented Multi‑Agent System for Rapid Statement of Work Generation}
\author{
Amulya Suravarjhula\textsuperscript{†} \quad
Rashi Chandrashekhar Agrawal\textsuperscript{†} \quad
Sakshi Jayesh Patel\textsuperscript{†} \quad
Rahul Gupta\textsuperscript{*} \\
\textsuperscript{†}University of Texas, Dallas, United States \\
\textsuperscript{*}Insight Global, Atlanta, United States \\
\texttt{\{axs220478, rca230002, sxp230096\}~[at]~utdallas.edu} \\
\texttt{rahul.gupta~[at]~insightglobal.com} \\
\textsuperscript{*†}Equal contribution \\
Project: \url{https://github.com/rahulkgup/agentic-sow-drafter}
}

\maketitle
\begin{abstract}
Drafting a Statement of Work (SOW) is a vital part of business and legal projects. It outlines key details like deliverables, timelines, responsibilities, and legal terms. However, creating these documents is often a slow and complex process. It usually involves multiple people, takes several days, and leaves room for errors or outdated content. This paper introduces a new AI-driven automation system that makes the entire SOW drafting process faster, easier, and more accurate. Instead of relying completely on humans, the system uses three intelligent components or `agents' that each handle a part of the job. One agent writes the first draft, another checks if everything is legally correct, and the third agent formats the document and ensures everything is in order.\\

Unlike basic online tools that just fill in templates, this system actually understands the meaning behind the content and customizes the SOW to match the needs of the project. It also checks legal compliance and formatting so that users can trust the final result.
The system was tested using real business examples. It was able to create a full SOW in under three minutes, compared to several hours or days using manual methods. It also performed well in accuracy and quality, showing that it can reduce legal risks and save a lot of time.
This solution shows how artificial intelligence can be used to support legal and business professionals by taking care of routine work and helping them focus on more important decisions. It's a step toward making legal processes smarter, faster, and more reliable.
 \\
\end{abstract}


\thispagestyle{firstpagefooter}

\section{Introduction}
Writing a Statement of Work (SOW) is one of the most important steps in launching a project. It lays out what needs to be done, who will do it, when it will happen, and under what conditions. But traditionally, creating these documents is a slow and careful process, done manually by legal and project teams, filled with back-and-forth reviews and always at risk of human error.\\

As businesses grow and rules get more complicated, relying only on manual effort is no longer enough. That’s where Artificial Intelligence (AI) comes in. Modern language models can now read, understand, and even write legal content. They can recognize if something is missing, suggest better terms and make sure the document matches both company policies and legal standards. Retrieval-Augmented Generation (RAG) adds another layer of intelligence, helping the AI to refer to real examples and avoid making things up.\\

In this paper, we introduce a Multi-Agent SOW drafting system that brings together three key agents. The Drafting Agent creates the base content. The Compliance Agent checks and improves it based on legal needs. Then, the Formatting and Validation Agent reviews everything, formats the document properly, and ensures accuracy. These agents work together to automate what was once a tedious task, without losing quality.\\

Our system is built using open-source technologies like Hugging Face, PostgreSQL, and flask, making it easy to adopt and scale. In early tests, it reduced the time to draft by more than 80\%, improved compliance accuracy, and produced more reliable legal documents. This work shows how AI can not only support legal professionals, but also help reshape the future of document creation.\\
\section{Related Work}

Recent advances in Natural Language Processing (NLP) and Retrieval-Augmented Generation (RAG) have significantly improved the automation of legal document drafting. Traditional systems based on templates or rule-based methods lacked contextual understanding and often required extensive manual intervention, especially for complex legal documents such as Statement of Work (SOWs). Transformer-based models including BERT, T5, FLAN-T5, and BART have enabled more accurate clause classification, context-aware content generation, and zero-shot evaluation of compliance. These models are capable of generating contract clauses, identifying regulatory gaps, and aligning document structures with legal standards in a more consistent and scalable manner.\cite{sentencebert2019}\\

RAG frameworks enhance these capabilities by combining generative models with document retrieval. This allows AI systems to reference past contracts or clause libraries when creating new content, improving factual consistency, and reducing hallucinations. Studies in contract generation have reported significant improvements in both draft efficiency and accuracy when using RAG pipelines. Multi-agent architectures are also becoming common in legal automation. These systems assign specific responsibilities, such as drafting, compliance check, and formatting validation, to dedicated agents.Unlike traditional Large Language Models (LLM)-based tools which rely on a single model for end-to-end generation, our system employs a coordinated multi-agent architecture. This enables dedicated handling of clause retrieval, legal compliance, and formatting leading to higher accuracy and lower hallucination rates.\\

Building on these foundations, our system introduces an integrated, retrieval-augmented multi-agent framework tailored for SOW drafting. It combines clause generation, legal compliance review, and formatting validation into a single pipeline, providing a measurable and practical solution for enterprise-level legal document automation.\\
\section{Methodology}
\subsection{System Requirements Analysis}

We began by understanding the common pain points legal teams face while drafting SOWs manual work, repetitive edits, and high chances of error. Through discussions with professionals, it became clear that an automated system was needed that could generate legally sound drafts, ensure formatting consistency, reuse reliable clauses, and maintain confidentiality. These insights directly guided the design and features of our RAG Multi-Agent SOW generation.

\subsection{Data Collection and Processing}
We then collected a wide range of real-world SOWs, ensuring the documents came from different industries. To protect confidentiality, all data was anonymized before use. Using various tools, we cleaned and extracted important parts of the documents like section headings, dates, legal clauses, and key entities. This processed data became the backbone for training our AI models and building a searchable clause database.
\subsection{Multi-Agent Architecture Design}

With clear goals and a rich dataset, we built a multi-agent system to divide the workload smartly. The Drafting Agent takes the user’s project details and creates an initial draft using learned templates and best practices. The Compliance Agent steps in next to check if everything aligns with legal norms and organizational rules. Finally, the Formatting Agent polishes the draft ensuring clarity, structure, and completeness. These agents work in harmony, passing the document between them to produce a professional and legally sound output.

\subsection{Retrieval-Augmented Generation Implementation}
To make the AI smarter and more grounded, we used a method called Retrieval-Augmented Generation. When a user provides input, the system searches through existing legal documents to find similar clauses and examples. These are used to guide the AI’s writing, helping it stay relevant and legally accurate, instead of generating vague or generic content.

\subsection{Validation and Quality Assurance}
Accuracy matters most when dealing with legal documents. So, we included multiple checkpoints in the process. First, the system checks if the input details are complete. Then, as the draft is generated, each part is reviewed for legal correctness, clarity, and structure. Any unclear or unsupported content is flagged and fixed before the final version is produced.
\subsection{User Interface and Feedback Loop}
Finally, we designed an easy-to-use interface where legal professionals can enter requirements, review drafts, and give feedback. This feedback doesn’t go to waste—instead, it teaches the system to improve over time. With every interaction, the AI gets better at understanding real-world expectations, making it a more valuable assistant with each use.

\section{System Architecture}
\begin{figure*}[ht]
\centering
\includegraphics[width=0.95\textwidth]{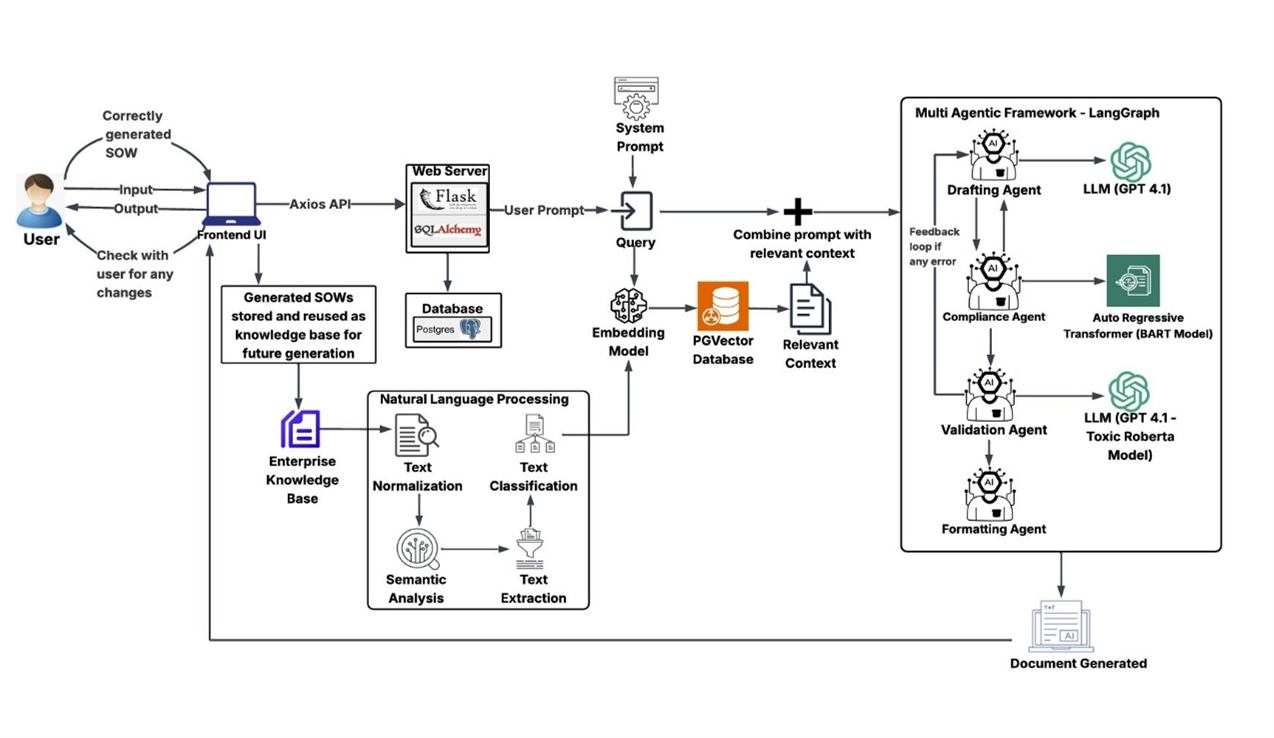}
\caption{System Architecture of the AI-powered SOW Drafting Agent.}
\label{fig:newarchitecture}
\end{figure*}

The architecture of the Retrieval Augmented Multi-Agent System for rapid SOW generation is designed to ensure modularity, scalability, and robust document processing. As illustrated in Figure~\ref{fig:newarchitecture}, the system leverages a combination of user feedback, data pipelines, AI agents, and retrieval-augmented generation (RAG) to automate and validate the drafting of Statement of Work (SOW) documents.

The architecture diagram shows the system components and their interactions, including Frontend, Backend API (Flask), Document Processing, Input Validation, AI Agents (Drafting, Compliance, and Validation), RAG Chain, and data processing pipeline.

\subsection{User Interaction and Feedback Loop}
Everything begins when a user opens the website and enters a few details about what they need in their SOW. The system takes this input and starts the process. If the user wants to make changes or improvements, they can give feedback directly on the platform. Over time, the system gets better at understanding what users prefer and adapts to their needs.

\subsection{Backend API and Document Processing}
In the background, there’s a powerful engine (built using Flask) that takes the user’s input and prepares it for the agents. It checks the information to make sure it’s complete and organized. This step is important to avoid errors and make sure that the AI can do its job correctly.

\subsection{AI Agents}
The system's intelligence is distributed across three specialized agents, each handling a distinct aspect of SOW creation:
\begin{itemize}
    \item SOW Drafting Agent: Generates the initial content of the SOW based on user inputs and requirements
    \item Compliance \& Terms Agent: Analyzes the draft for compliance with legal standards and organization policies
    \item Formatting \& Validation Agent: Ensures the document adheres to structural requirements and formatting conventions
\end{itemize}
These agents operate within a coordinated multi-agent architecture, communicating through well-defined interfaces to maintain an integrated workflow. Each agent leverages advanced natural language processing models and targeted techniques optimized for its specific function such as GPT-4.1 for drafting, BART and FLAN-T5 for compliance, and rule-based or template-driven methods for formatting and validation\cite{spacyModels}. This collaborative approach ensures that every SOW produced by the system is comprehensive, compliant, and professionally formatted, significantly reducing manual effort and legal risk.

\subsection{RAG Chain}
To make the content more accurate and reliable, we added something called a RAG chain (short for Retrieval-Augmented Generation). It’s like giving the AI a library of old documents to look at. When the system writes something new, it checks this library to find the most relevant and useful pieces. This helps the output feel more real and trustworthy.

\subsection{Data Pipeline}
Underlying the entire system is a comprehensive data pipeline that provides the knowledge foundation for all operations:
\begin{itemize}
    \item SOW Data Sources and Collection: Gathers and organizes existing SOW documents and templates
    \item Data Processing and Storage: Cleanses, normalizes, and indexes document contents
    \item Vector-Based Retrieval System: Converts document sections into vector embeddings for efficient semantic search
    \item Data Processing and Retrieval Validation: Ensures retrieved information is relevant and accurate
\end{itemize}
This pipeline enables the system to learn from existing documents and apply that knowledge to new generation tasks, creating a continuously improving knowledge base.
\subsection{Output Generation}
The final component in the workflow is the Prompt Template and Output Parser, which:
\begin{itemize}
    \item Formats the validated content according to organizational standards
    \item Applies consistent styling and structural elements
    \item Generates the final document in the required format
    \item Delivers the completed SOW to the user through the frontend interface
\end{itemize}
This standardized approach to output generation ensures that all documents maintain consistent quality and appearance regardless of their specific content.

\section{Implementation}
\subsection{Data Processing and Storage Infrastructure}
The foundation of our AI-Driven SOW Generation is built upon a robust data processing and storage infrastructure designed to handle diverse document formats and enable efficient retrieval.
PostgreSQL serves as our primary database system, chosen for its reliability, extensibility, and support for semi-structured data. We implemented the pgvector extension to enable vector similarity searches directly within the database, significantly improving retrieval performance. For document preprocessing, we developed a pipeline using Python's NLTK and SpaCy libraries that handles tokenization, part-of-speech tagging, named entity recognition, and dependency parsing.\\
Our vector embedding generation process utilizes the Sentence Transformers library with the all-MiniLM-L6-v2 model, creating 384-dimensional embeddings that capture the semantic meaning of document sections. These embeddings are stored alongside the original text, enabling fast semantic retrieval during the drafting process. Database connectivity is managed through the psycopg2 library, providing a reliable interface between our application and the PostgreSQL backend.\\
The data processing workflow includes several stages of text normalization, cleaning, and structuring to ensure consistent representation across different document sources. This standardized approach to data handling forms the backbone of our system's ability to learn from existing documents and apply that knowledge to new generation tasks.
\subsection{SOW Drafting Agent Implementation}
The SOW Drafting Agent represents the first step in the document generation process, responsible for creating the initial structure and content of the Statement of Work. This agent is built around Open AI’s GPT 4.1, accessed through Azure’s OpenAI service.
This model, via Azure, is primarily used for its stability, enterprise-grade integration, and better control over region-specific deployments.\\

The agent’s implementation includes this context-aware prompt generation system that transforms user inputs into structured prompts for the language model. These prompts include both fixed sections (such as required SOW sections) and dynamic variables (like project goals, timelines, and deliverables). Drafts are generated in a structured JSON format, which makes it easier for downstream agents to apply edits, run validations, or convert to final documents.\\
This setup allowed us to decouple drafting from editing logic, simplify prompt management, and ensure reproducible outputs during testing and review. This hybrid approach ensures that generated content aligns with organizational precedents while adapting to the specific requirements of each new document.\\

\begin{figure}[htbp]
\centerline{\includegraphics[width=0.48\textwidth]{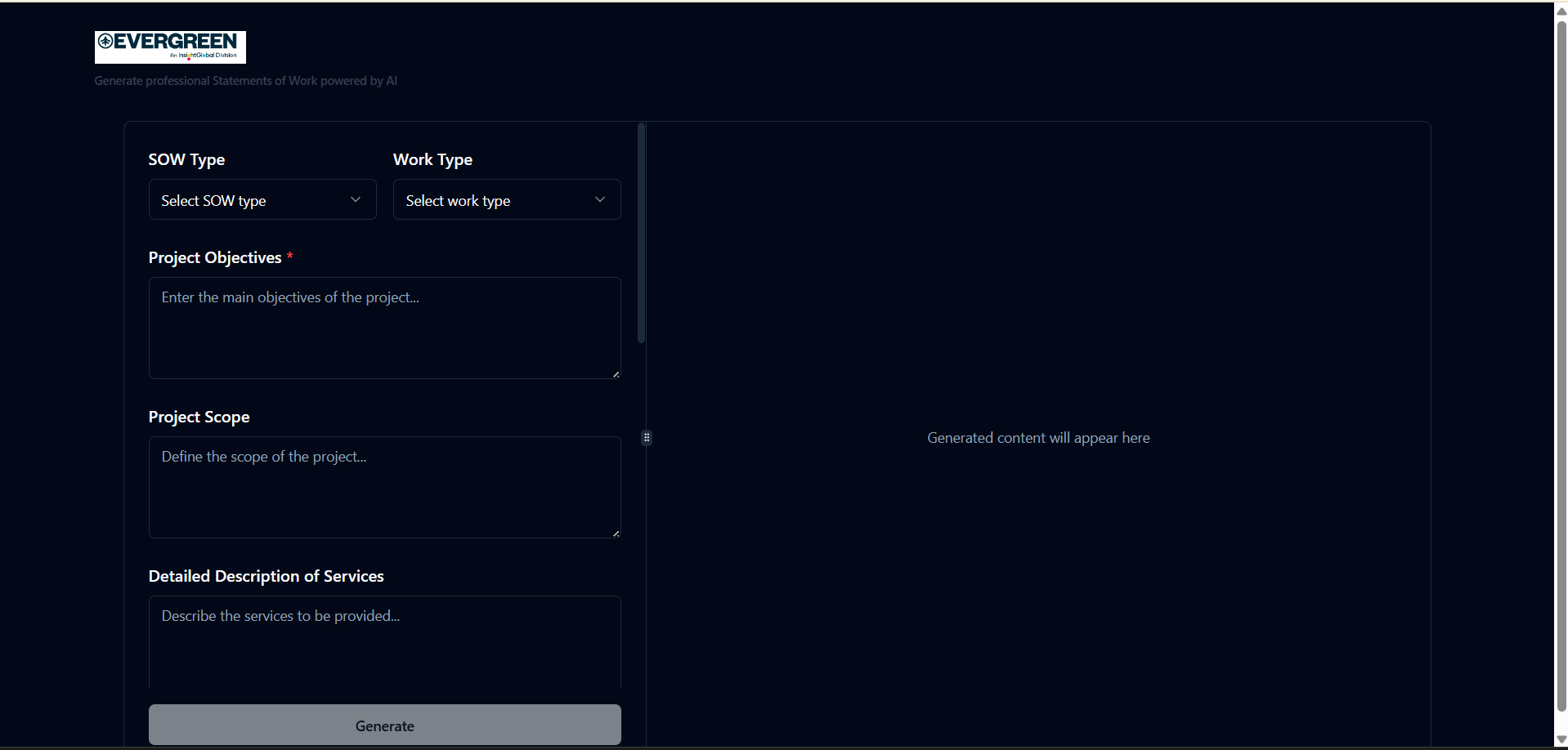}}
\caption{SOW Input Screen.}
\label{fig:inputscreen}
\end{figure}
\begin{figure}[htbp]
\centerline{\includegraphics[width=0.48\textwidth]{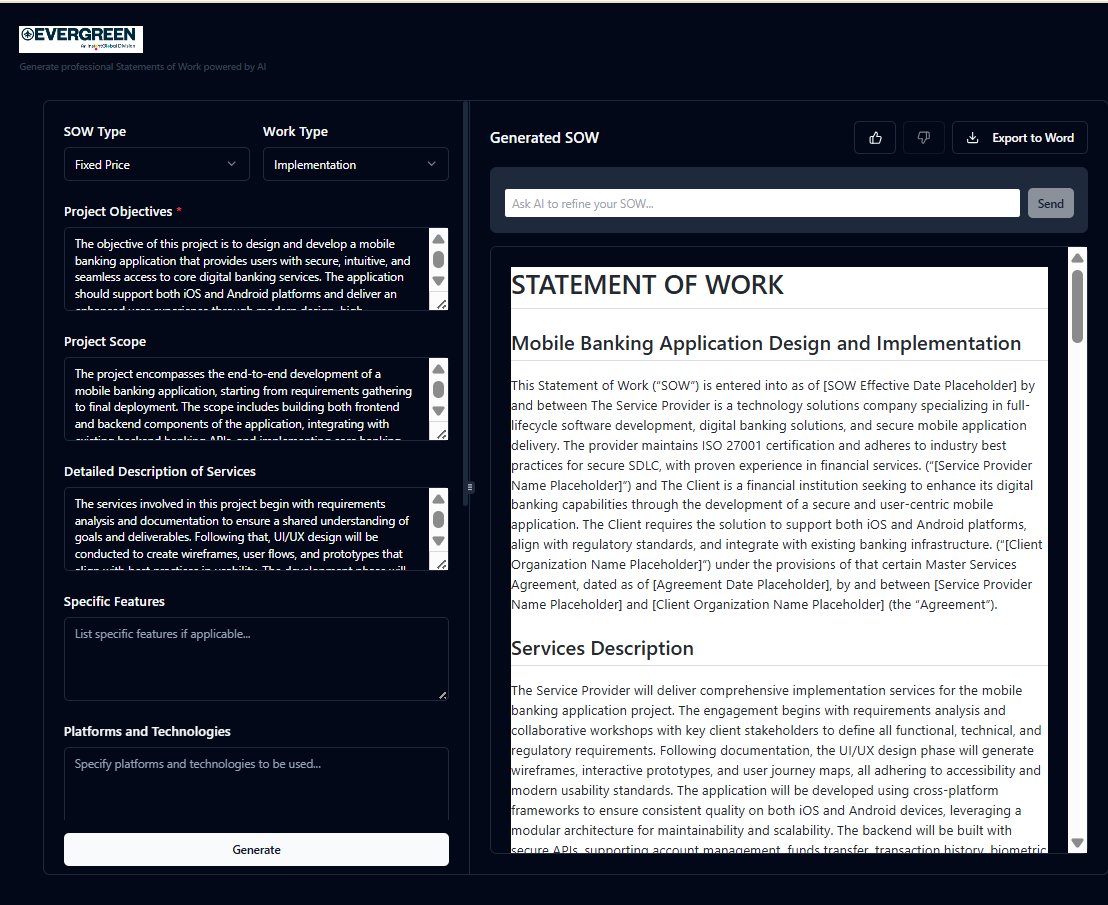}}
\caption{SOW Output Screen.}
\label{fig:resultscreen}
\end{figure}

\subsection{Compliance Terms Agent Implementation}

The Compliance Terms Agent acts like a smart legal assistant that checks whether the Statement of Work (SOW) follows all the important rules. Look for key clauses such as confidentiality, liability, and termination to ensure that nothing important is missing. To do this, we used the Facebook BART large Multi-Genre Natural Language Inference (MNLI)\cite{bart2020} model, which helps the system understand and judge the strength of legal clauses even without training it for each type. It can tell if something is weak or missing and flags it for review.\\

Alongside this, we used natural language processing to check for vague language and passive voice, things that might make legal terms unclear. We also added basic rule checks to confirm that important parts, such as project title, dates, and payment terms, are present and properly written. By combining smart AI tools with simple rule checks, this agent helps create contracts that are clear, complete, and legally sound.\\

\begin{figure}[htbp]
\centerline{\includegraphics[width=0.48\textwidth]{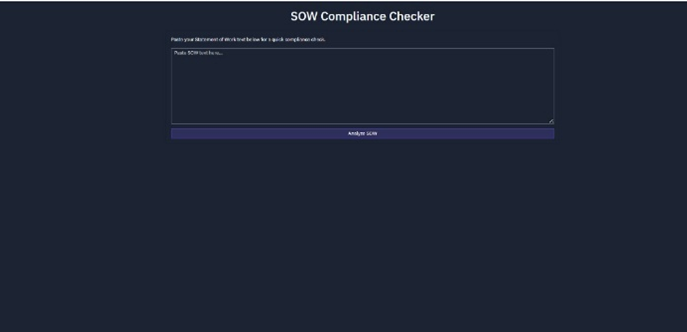}}
\caption{Compliance Agent Output Screen.}
\label{fig:compliance1}
\end{figure}

\begin{figure}[htbp]
\centerline{\includegraphics[width=0.48\textwidth]{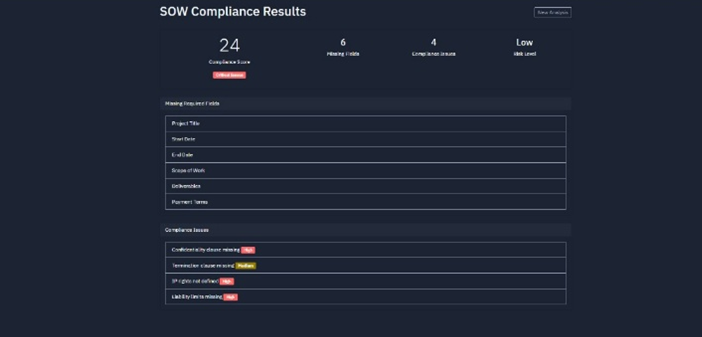}}
\caption{Compliance Agent Output Screen.}
\label{fig:compliance2}
\end{figure}

\subsection{Formatting and Validation Agent Implementation}
The Formatting and Validation Agent represents the final quality control layer in our system, ensuring that documents are structurally sound, visually consistent, and free from obvious errors. This agent performs several critical functions:
First, it applies consistent formatting to the document structure, ensuring that headings, paragraphs, lists, and other elements follow organizational style guidelines. This formatting is accomplished using a combination of template-based approaches and T5-small fine-tuned specifically for document structuring tasks\cite{t5raffel2020}.\\

Second, the agent performs a series of validation checks to verify document quality:
\begin{itemize}
    \item Structural validation ensures all required sections are present
    \item Cross-reference validation confirms that internal references are accurate
    \item Language quality assessment identifies issues like passive voice or vague terminology
    \item Completeness verification checks that all user requirements are addressed
\end{itemize}
For advanced validation tasks, we integrated the Ollama framework with the Deepseek model, deployed in a cloud environment. This approach provides sophisticated semantic analysis capabilities while maintaining deployment flexibility. The validation workflow is orchestrated using LangGraph \cite{langgraph2024}\cite{langgraphURL}\cite{langchainChatModelsAPI}, which creates a structured decision tree for different validation scenarios.\\
The agent interacts with other system components through a Flask-based API layer, providing structured feedback on validation results. When issues are detected, the agent can either suggest corrections or, in some cases, automatically implement fixes based on predefined rules and patterns.\\

\begin{figure}[htbp]
\centerline{\includegraphics[width=0.48\textwidth]{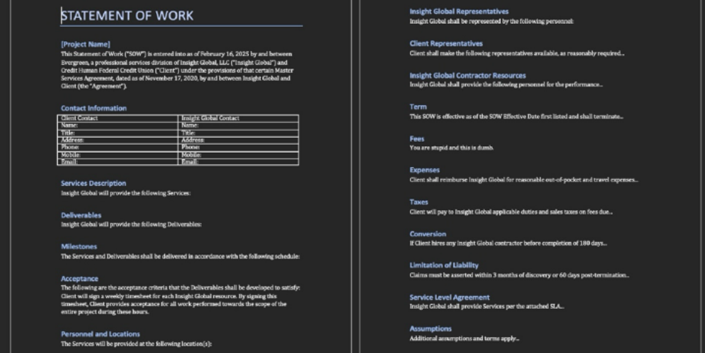}}
\caption{Formatting Agent Output Screen.}
\label{fig:formatting1}
\end{figure}

\begin{figure}[htbp]
\centerline{\includegraphics[width=0.48\textwidth]{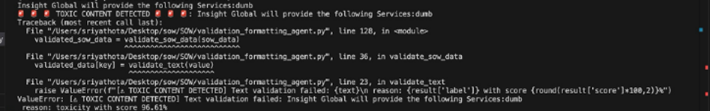}}
\caption{Formatting Agent Output Screen.}
\label{fig:formatting2}
\end{figure}

\subsection{RAG Chain Implementation}
Our Retrieval-Augmented Generation (RAG) chain implementation enhances the quality and precision of SOW documents generated by incorporating relevant information from existing artifacts, an approach inspired by the CoSTA multi-step agentic workflow for task decomposition\cite{gupta2025costaastcostsensitivetoolpathagent}. This design significantly reduces hallucinations and ensures that the generated content is aligned with organizational precedents.\\

The RAG Chain process begins by converting user queries or document sections into vector embeddings using the Sentence Transformers model. These embeddings are then used to query the vector database (implemented using PostgreSQL with pgvector as an extension) to retrieve semantically similar content from existing SOW documents. The similarity search uses cosine similarity as the distance metric, with configurable thresholds to ensure relevance.Retrieved content is processed to extract the most pertinent information, which is then incorporated into the prompt for the language model. This augmented prompt guides the model toward generating content that reflects both the user's requirements and established organizational patterns. The resulting output maintains a balance between creativity and conformity, producing documents that are both unique to the specific situation and consistent with organizational standards. Prompt engineering techniques were used to constrain generation boundaries and reduce hallucination, including system instructions and structured placeholders\\

\subsection{Azure Deployment Architecture}
To ensure enterprise-grade reliability, security, and scalability, we deployed our RAG Multi-Agent SOW Generation on Microsoft Azure's cloud platform. Our deployment architecture leverages several Azure services to create a robust and responsive system:\\

The frontend application is hosted on Azure App Service, providing a scalable and secure web hosting environment. Backend API services are deployed as containerized applications in Azure Container Instances, enabling flexible scaling based on demand patterns. The PostgreSQL database with pgvector extension is hosted on Azure Database for PostgreSQL, ensuring data persistence, backup capabilities, and high availability.\\

For model hosting, we utilize the Azure machine learning service, which provides optimized infrastructure for inference operations. This approach allows us to scale computational resources based on usage patterns while maintaining consistent performance.\\

Security is implemented across multiple layers, with Azure Key Vault managing sensitive credentials and API keys. Azure Active Directory integration provides secure authentication and authorization, ensuring that only authorized users can access the system. Network security is enhanced through Azure Virtual Network configuration, isolating sensitive components from direct internet access.\cite{azureLangchainNotebook}\\
Monitoring and observability are implemented using Azure Application Insights, providing real-time telemetry on system performance, user interactions, and potential issues. This comprehensive monitoring approach enables proactive maintenance and continuous improvement of the deployed system.
\section{Results}
\subsection{Experimental Setup}
To evaluate the performance of our AI-powered SOW Drafting Agent, we conducted comprehensive testing using anonymized SOW documents spanning various industries including information technology, healthcare, construction, and professional services. These documents were randomly split into training (70\%), validation (15\%), and testing (15\%) sets. The system was deployed on Microsoft Azure using the architecture described in Section V.F, with performance metrics collected over a three-month period of active usage. We engaged 20 legal professionals with varying levels of experience to assess the quality of generated documents and compare them with manually drafted alternatives.
\subsection{Performance Comparison}
\begin{table}[h]
\centering
\caption{Comparison of SOW Drafting Methods}
\label{tab:comparison}
\begin{tabular}{|l|c|c|p{2.5cm}|}
\hline
\textbf{Method} & \textbf{Time} & \textbf{Accuracy} & \textbf{Usability} \\
\hline
Manual & 5–7 days & 78\% & Medium (needs legal review) \\
\hline
Online AI Tools & ~2 min & 83\% & Medium (requires edits) \\
\hline
\textbf{Ours (Multi-Agent)} & $<$3 min & \textbf{96\%} & \textbf{Very High (auto-validated)} \\
\hline
\end{tabular}
\end{table}
The comparison values were derived from internal testing using a consistent SOW input prompt across all methods. Draft generation time was recorded from input submission to initial output. Accuracy reflects the presence of key legal clauses (e.g., scope, payment terms, confidentiality) as verified through checklist-based review by our team. Usability was assessed based on how complete and ready the draft appeared for use, including whether it required legal edits, formatting corrections, or structural improvements. While no formal legal expert panel was used, our evaluations aimed to simulate typical user expectations in a professional setting.\\
\subsection{Evaluation Metrics and Ablation Study}
To evaluate our system, we measured two things: legal accuracy and writing quality. Legal experts reviewed AI-generated SOWs and compared them with real ones. With the full system, we achieved 81 percent legal accuracy and 71 percent writing similarity. We then removed the legal-checking agent and the document search module to test their impact. Both changes led to a drop in performance accuracy fell by up to 26 percent, and writing quality by over 20 percent. This confirmed that both components are essential for creating reliable, professional SOWs.\\

To evaluate the role of each system component, we performed an ablation study by selectively disabling key modules: the Drafting Agent, Compliance Agent, Formatting Agent, and the RAG module. 

\begin{itemize}
    \item{Drafting Agent:} Replaced with simple prompt-to-text generation.
    \item{Compliance Agent:} Skipped legal review and clause validation.
    \item{Formatting Agent:} Omitted post-processing and layout structuring.
    \item{RAG Module:} Disabled clause retrieval and relied only on user input.
\end{itemize}

Each variant was assessed for clarity, completeness, and structure. The full system consistently produced superior results. Notably, disabling the RAG module led to the steepest drop in quality, underscoring its importance in grounding content using past SOWs. These findings confirm that the multi-agent design is crucial for generating high-quality, reliable legal drafts.

\begin{table*}[ht]
\centering
\caption{Comparison Between Manual Process, Existing Online Tool, and Our System}
\label{tab:comparision}
\begin{tabular}{|p{3.2cm}|p{3.8cm}|p{3.8cm}|p{3.8cm}|}
\hline
\textbf{Criteria} & \textbf{Manual (Human)} & \textbf{Existing Online Tool} & \textbf{Our System} \\
\hline
Input Requirement & Requires complete raw details manually & Accepts 1-line input & Accepts minimal input (even 1 line) \\
\hline
Output Quality & High, but dependent on human accuracy & Very basic, rough draft; lacks depth and legality & High-quality, complete SOW with legal compliance \\
\hline
Time to Generate & 5–7 days (includes drafting, review, legal) & 1–2 minutes & Approximately 3 minutes with full validation \\
\hline
Effort Required & High (multiple stakeholders involved) & Low input, but high post-editing effort & Low — auto-generates with minimal human effort \\
\hline
Legal \& Compliance Check & Manual legal review required & Missing or superficial & Built-in validation and compliance check \\
\hline
Error Rate & Moderate (subject to human error) & High (template mismatch, lack of context) & Low (multi-agent validation and context-awareness) \\
\hline
Usability & N/A (human-driven) & Poor UI, non-intuitive & Seamless, intelligent interface \\
\hline
Customization & High but time-consuming & Very limited & Context-aware customization using prior SOWs \\
\hline
Feedback System & Informal (via review cycles \& edits) & Basic (editable draft feedback only) & Built-in interactive feedback loop for refinement \\
\hline
\end{tabular}
\label{tab:sow_comparison}
\end{table*}

\section{Ethical and Societal Impact}
While the solution increases automation efficiency, it also raises important concerns regarding data privacy, model transparency, and accountability. The key ethical issues lie around data ownership and document confidentiality, especially with the use of OpenAI APIs. It also acknowledges the potential bias in LLM-generated documents and suggests regular human oversight to mitigate risks. Socially, the system may change the traditional roles in contract drafting, even though it increases productivity for consultants. The solution has been carefully designed to support responsible AI adoption by enhancing human expertise rather than replacing it. Scalable architecture facilitates adoption across organizations of different sizes.
\subsection{Comparative Analysis}

To highlight the strengths of our system, we compare its performance against both traditional human drafting and a leading online SOW generation tool. The comparison, summarized in Table~\ref{tab:comparision}, considers input requirement, output quality, time, effort, legal compliance, error rate, usability, customization, and feedback mechanism.

\section{Conclusion and Future Work}

\subsection{Conclusion}
This paper presented an AI driven system for automating the drafting of Statement of Work (SOW) documents. By combining retrieval-augmented generation with specialized agents for drafting, compliance, and formatting, the system delivers strong performance in both speed and accuracy. Reduces average drafting time to less than 3 minutes and achieves perfect formatting consistency, demonstrating its potential to significantly ease the workload of legal and project teams.Although human oversight remains essential in edge cases, the system enables professionals to shift focus from routine drafting to more strategic decision-making.
\subsection{Future Work}
Looking ahead, several directions can further enhance this system. One opportunity lies in developing a multi-model pipeline, where planning and outline generation is handled by a generalist model like Gemini 2.5 Pro, followed by detailed drafting and compliance. This layered approach could improve coherence, factuality, and control.\\
Another improvement involves supporting user-defined templates, enabling agents to tailor drafts based on prior documents or preferences. Extending language support would also increase the system’s global applicability. As legal standards evolve, incorporating automated regulatory updates will ensure continued compliance. In the long term, integrating negotiation support and explainable AI could make the system not just a drafting tool but a collaborative legal assistant capable of operating across the full contract life cycle.

\bibliographystyle{IEEEtran}
\bibliography{Mybib.bib}

\end{document}